\documentclass[preprintnumbers,amsmath,amssymbm,prd]{revtex4}
\usepackage{epsfig}
\usepackage{graphicx}
\usepackage{amssymb}

\begin{document}
\title{Energy spectrum of the long-range Lennard-Jones potential}
\author{Shahar Hod}
\affiliation{The Ruppin Academic Center, Emeq Hefer 40250, Israel}
\affiliation{ } \affiliation{The Hadassah Institute, Jerusalem
91010, Israel}
\date{\today}

\begin{abstract}
\ \ \ The discrete energy spectra of composite inverse power-law
binding potentials of the form
$V(r;\alpha,\beta,n)=-\alpha/r^2+\beta/r^n$ with $n>2$ are studied
{\it analytically}. In particular, using a functional matching
procedure for the eigenfunctions of the radial Schr\"odinger
equation, we derive a remarkably compact analytical formula for the
discrete spectra of binding energies
$\{E(\alpha,\beta,n;k)\}^{k=\infty}_{k=1}$ which characterize the
highly-excited bound-state resonances of these long-range binding
potentials. Our results are of practical importance for the physics
of polarized molecules, the physics of composite polymers, and also
for physical models describing the quantum interactions of bosonic
particles.
\end{abstract}
\bigskip
\maketitle


\section{Introduction}

Inverse power-law potentials are ubiquitous in physics and chemistry. In particular, they play a central role in physical models describing the interactions of atoms and molecules (see \cite{NC1,NC2,NC3,NC4,NC5,NC6,NC7,NC8,NC9,NC10,LanLif} and references therein). Of particular interest is the effective Lennard-Jones potential \cite{LJ}
\begin{equation}\label{Eq1}
V(r;\alpha,\beta,n_1,n_2)_{\text{LJ}}=-{{\alpha}\over{r^{n_1}}}+{{\beta}\over{r^{n_2}}}\ \ \ \ \text{with}\ \ \ \ n_2>n_1>0\ ,
\end{equation}
which is composed of two inverse power-law interaction terms: the negative term $-\alpha/r^{n_1}$ describes an attractive interaction between the components of the system, whereas the positive term $\beta/r^{n_2}$ describes an effective repulsive interaction \cite{NotePa}. The standard form of the Lennard-Jones potential is characterized by the inequality $n_2>n_1$, in which case the radial function (\ref{Eq1}) has the form of an effective potential well whose minimum is located at $r^{\text{min}}=(\beta n_2/\alpha n_1)^{{1}\over{n_2-n_1}}$.

The most studied form of the Lennard-Jones binding potential (\ref{Eq1}) is known as the 6-12 Lennard-Jones potential, which is characterized by an attractive Van der Waals interaction term with $n_1=6$. In this case (that is, $n_1>2$) one finds that the effective potential well, which is described by the composite inverse power-law functional relation (\ref{Eq1}), can only support a {\it finite} number of bound-state resonances \cite{LanLif}.

On the other hand, there are known quantum mechanical systems which are characterized by
attractive long-range inverse square potentials with $n_1=2$ \cite{Sq1,Sq2,Sq3,Sq4,Sq5,Sq6,Sq7,Sq8,Sq9,Sq10}.
For instance, the inverse square law potential naturally appears in physical models describing the
interactions between polar molecules and laser-excited Rydberg atoms \cite{Sq3,Sq4,Sq5,Sq6},
in the Efimov effect describing the interactions of bosonic particles \cite{Sq7,Sq8}, and also
in theoretical models describing the physics of composite polymers \cite{Sq9,Sq10}.
The attractive inverse square potential is especially interesting from physical as well
as mathematical points of view since it sits on the boundary between long-range power-law
potentials with $n_1<2$, which can support an infinite number of bound-state resonances \cite{LanLif},
and short-range power-law potentials with $n_1>2$ which can only support a finite number of bound-state
resonances \cite{LanLif}.

The main goal of the present paper is to explore the physical
properties of effective binding potentials of the Lennard-Jones form
(\ref{Eq1}) with an attractive {\it long}-range inverse square law
(that is, $n_1=2$) interaction
term\cite{Sq1,Sq2,Sq3,Sq4,Sq5,Sq6,Sq7,Sq8,Sq9,Sq10}. In particular,
we shall analyze the discrete energy spectra
$\{E(\alpha,\beta,n;k)\}^{k=\infty}_{k=1}$ \cite{Notekr} which
characterize the family \cite{NoteLR1,NoteLR2,Notelll,Notespt}
\begin{equation}\label{Eq2}
V(r;\alpha,\beta,n)_{\text{LR-LJ}}=-{{\alpha}\over{r^{2}}}+{{\beta}\over{r^{n}}}\ \ \ \ \text{with}\ \ \ \ n>2\
\end{equation}
of composed long-range inverse power-law binding potentials.

Interestingly, below we shall explicitly show that the Schr\"odinger differential equation with the effective radial potentials (\ref{Eq2}) is amenable to an analytical treatment in the regime of small binding energies. In particular, we shall derive a compact {\it analytical} formula [see Eq. (\ref{Eq36}) below] for the discrete energy spectra $\{E(\alpha,\beta,n;k)\}^{k=\infty}_{k=1}$ which characterize the highly-excited bound-state resonances of the composed {\it long}-range Lennard-Jones potentials (\ref{Eq2}).

\section{Description of the system}

We shall study the stationary bound-state ($E<0$) resonances of the Schr\"odinger differential equation \cite{Notemul}
\begin{equation}\label{Eq3}
\Big[-{{\hbar^2}\over{2\mu}}{{d^2}\over{dr^2}}+V(r)_{\text{LR-LJ}}\Big]\psi=E \psi\  ,
\end{equation}
where the spatial behavior of the binding potential $V(r)_{\text{LR-LJ}}$ is described by the composed inverse power-law (long-range Lennard-Jones) functional relation (\ref{Eq2}). The stationary eigenstates of the radial Schr\"odinger equation (\ref{Eq3}) are characterized by spatially bounded radial eigenfunctions:
\begin{equation}\label{Eq4}
\psi(r\to\infty)\sim e^{-\omega r}\  ,
\end{equation}
where the characteristic frequency is defined by the relation \cite{Notew0}
\begin{equation}\label{Eq5}
\omega^2\equiv -{{2\mu}\over{\hbar^2}}E\ \ \ \ \text{with}\ \ \ \ \omega\in \mathbb{R}\  .
\end{equation}
In addition, we shall look for bound-state radial eigenfunctions which are spatially regular at the origin:
\begin{equation}\label{Eq6}
\psi(r\to0)<\infty\  .
\end{equation}

The Schr\"odinger differential equation (\ref{Eq3}), together with the physically motivated boundary conditions
(\ref{Eq4}) and (\ref{Eq6}), determine the discrete spectra of binding energies $\{E(\alpha,r_0,n)\}$ which characterize the composed inverse power law binding potentials (\ref{Eq2}). As we shall explicitly show below, the qualitative behavior of the quantum system (\ref{Eq3}) is determined by the characteristic length-scale
\begin{equation}\label{Eq7}
r_0\equiv \Big({{2\mu\beta}\over{\hbar^2}}\Big)^{1/(n-2)}\  .
\end{equation}
In particular, in the next section we shall use a functional matching procedure for the characteristic eigenfunctions
of the radial Schr\"odinger equation (\ref{Eq3}) in order to show that, in the regime \cite{Notese}
\begin{equation}\label{Eq8}
\omega r_0\ll1\
\end{equation}
of weakly bound-state resonances, the discrete energy spectra $\{E(\alpha,r_0,n)\}$ of the composite inverse power law binding potentials (\ref{Eq2}) [the long-range Lennard-Jones potentials (\ref{Eq2})] can be determined {\it analytically}.

\section{Asymptotic analytical solutions of the radial Schr\"odinger equation}

In the present section we shall analyze the Schr\"odinger differential equation
\begin{equation}\label{Eq9}
\Big\{{{d^2}\over{dr^2}}-{{1}\over{r^2}}\Big[(\omega r)^2-({1\over4}+\nu^2)+\Big({{r_0}\over{r}}\Big)^{n-2}\Big]\Big\}\psi(r;\omega,\nu,r_0,n)=0\
\end{equation}
for the long-range Lennard-Jones \cite{NoteLR2} binding potentials (\ref{Eq2}), where \cite{Notenu0}
\begin{equation}\label{Eq10}
{1\over4}+\nu^2\equiv {{2\mu\alpha}\over{\hbar^2}}\ \ \ ; \ \ \ \nu\in\mathbb{R}\  .
\end{equation}
In particular, we shall explicitly show below that
the radial Schr\"odinger equation (\ref{Eq9}), which determines the energy spectra of the composite inverse power-law potentials (\ref{Eq2}), is amenable to an analytical treatment
in the asymptotic radial regions $r\ll 1/\omega$ and $r\gg r_0$. We shall then show that, by using a functional matching procedure for the characteristic radial solutions of (\ref{Eq9}) in the overlapping radial region $r_0\ll r\ll 1/\omega$, one can determine {\it analytically} the discrete spectra of eigen-frequencies $\{\omega(\nu,r_0,n)\}$ which characterize the weakly-bound (highly-excited) resonances of the  long-range Lennard-Jones potentials (\ref{Eq2}) \cite{NoteLR2}.

We shall first obtain an analytical expression for the radial eigenfunction $\psi(r)$ in the radial region
\begin{equation}\label{Eq11}
r\ll 1/\omega\  ,
\end{equation}
in which case one may approximate the Schr\"odinger differential equation (\ref{Eq9}) by
\begin{equation}\label{Eq12}
\Big({{d^2}\over{dr^2}}+{{{1\over4}+\nu^2}\over{r^2}}-{{r^{n-2}_0}\over{r^n}}\Big)\psi=0\  .
\end{equation}
The general mathematical solution of the differential equation (\ref{Eq12}) can be expressed in terms of the Hankel functions of the first and second kinds (see Eq. 9.1.53 of \cite{Abram}):
\begin{equation}\label{Eq13}
\psi(r)=A_1 r^{1\over
2}H^{(1)}_{{2{\bar\nu}}\over{n-2}}\Big[{{2i}\over{n-2}}\Big({{r_0}\over{r}}\Big)^{(n-2)/2}\Big]+
A_2 r^{1\over
2}H^{(2)}_{{2{\bar\nu}}\over{n-2}}\Big[{{2i}\over{n-2}}\Big({{r_0}\over{r}}\Big)^{(n-2)/2}\Big]
\ ,
\end{equation}
where $\{A_1,A_2\}$ are normalization constants and ${\bar\nu}\equiv
i\nu$.

Using the large-argument ($r_0/r\to\infty$) asymptotic behaviors
\begin{equation}\label{Eq14}
H^{(1)}_{a}(x\to\infty)={{(1-i)e^{-ia\pi/2}}\over{\sqrt{\pi}}}{{e^{ix}}\over{\sqrt{x}}}\cdot[1+O(x^{-1})]\ \ \ ; \ \ \ H^{(2)}_{a}(x\to\infty)={{(1+i)e^{ia\pi/2}}\over{\sqrt{\pi}}}{{e^{-ix}}\over{\sqrt{x}}}\cdot[1+O(x^{-1})]\
\end{equation}
of the Hankel functions, one finds from (\ref{Eq13}) the limiting behavior
\begin{equation}\label{Eq15}
\psi(r\to0)=r^{1\over2}_0\Big({{n-2}\over{\pi}}\Big)^{1\over2}\Big({{r}\over{r_0}}\Big)^{n\over4}\cdot\Big\{-iA_1e^{{{\nu\pi}\over{n-2}}}\exp\Big[-{{2}\over{n-2}}\Big({{r_0}\over{r}}\Big)^{(n-2)/2}\Big] +A_2e^{-{{\nu\pi}\over{n-2}}}\exp\Big[{{2}\over{n-2}}\Big({{r_0}\over{r}}\Big)^{(n-2)/2}\Big]\Big\}\
\end{equation}
for the radial eigenfunctions near the origin. The physically acceptable bound-state resonances of the binding potentials (\ref{Eq2}) are characterized by regular radial eigenfunctions at the origin $r\to0$ [see the boundary condition (\ref{Eq6})]. One therefore concludes that the coefficient of the exploding exponent in (\ref{Eq15}) should vanish:
\begin{equation}\label{Eq16}
A_2=0\  .
\end{equation}

Using the small-argument ($r_0/r\ll1$) asymptotic behavior
\begin{equation}\label{Eq17}
H^{(1)}_{a}(x\to0)=\Big\{\Big[{{1}\over{\Gamma(1+a)}}-i{{\cos(a\pi)\Gamma(-a)}\over{\pi}}\Big]\cdot(x/2)^{a}-i{{\Gamma(a)}\over{\pi}}\cdot(x/2)^{-a}\Big\}\cdot[1+O(x^2)]\
\end{equation}
of the Hankel function, one finds from (\ref{Eq13}) the power-law behavior
\begin{eqnarray}\label{Eq18}
\psi(r)=A_1
\Bigg\{{r^{1\over2}_0\Big({{i}\over{n-2}}\Big)^{{2{\bar\nu}}\over{n-2}}}\Bigg[{{1}\over{\big({{2{\bar\nu}}
\over{n-2}}\big)\Gamma\big({{2{\bar\nu}}\over{n-2}}\big)}}-i{{\cos\big({{2{\bar\nu}}\over{n-2}}\pi\big)
\Gamma\big(-{{2{\bar\nu}}\over{n-2}}\big)}\over{\pi}}\Bigg]\cdot\Big({{r}\over{r_0}}\Big)^{{1\over2}-{\bar\nu}}
-ir^{1\over2}_0{{\Gamma\big({{2{\bar\nu}}\over{n-2}}\big)}\over{\pi\big({{i}\over{n-2}}\big)^{{2{\bar\nu}}
\over{n-2}}}}\cdot\Big({{r}\over{r_0}}\Big)^{{1\over2}+{\bar\nu}}\Bigg\}\
\end{eqnarray}
for the characteristic radial eigenfunctions of the
Schr\"odinger differential equation (\ref{Eq9}) in the intermediate radial region
\begin{equation}\label{Eq19}
r_0\ll r \ll 1/\omega\  .
\end{equation}

We shall next obtain an analytical expression for the radial eigenfunction $\psi(r)$ in the radial region
\begin{equation}\label{Eq20}
r\gg r_0\  ,
\end{equation}
in which case one may approximate the Schr\"odinger differential
equation (\ref{Eq9}) by
\begin{equation}\label{Eq21}
\Big({{d^2}\over{dr^2}}-\omega^2+{{{{1\over4}+\nu^2}}\over{r^2}}\Big)\psi=0\ .
\end{equation}
The general mathematical solution of the differential equation
(\ref{Eq21}) can be expressed in terms of the Hankel functions of
the first and second kinds (see Eq. 9.1.49 of \cite{Abram})
\begin{equation}\label{Eq22}
\psi(r)=B_1r^{1\over 2}H^{(1)}_{{\bar\nu}}({\bar\omega}
r)+B_2r^{1\over 2}H^{(2)}_{{\bar\nu}}({\bar\omega} r)\ ,
\end{equation}
where $\{B_1,B_2\}$ is a normalization constants and
${\bar\omega}\equiv i\omega$.

Using the large-argument ($\omega r\to\infty$) asymptotic behaviors (\ref{Eq14}) of the Hankel functions, one finds from (\ref{Eq22}) the asymptotic behavior
\begin{equation}\label{Eq23}
\psi(r\to\infty)=B_1{{-\sqrt{2}ie^{\nu\pi/2}}\over{\sqrt{\omega\pi}}}e^{-\omega r}+B_2{{\sqrt{2}e^{-\nu\pi/2}}\over{\sqrt{\omega\pi}}}e^{\omega r}
\end{equation}
for the radial eigenfunctions. The physically acceptable bound-state resonances of the effective radial potentials (\ref{Eq2}) are characterized by finite (exponentially decaying) radial eigenfunctions at spatial infinity $r\to\infty$ [see the boundary condition (\ref{Eq4})]. One therefore concludes that the coefficient of the exploding exponent in (\ref{Eq23}) should vanish \cite{Noterc}:
\begin{equation}\label{Eq24}
B_2=0\  .
\end{equation}

Using the small-argument ($\omega r\ll1$) asymptotic behavior (\ref{Eq17}) of the Hankel
function, one finds from (\ref{Eq22}) the power-law behavior
\begin{eqnarray}\label{Eq25}
\psi(r)=B_1\Big\{{({\bar\omega}/2)^{{\bar\nu}}}\Big[{{1}
\over{{\bar\nu}\Gamma({\bar\nu})}}-i{{\cos({\bar\nu}\pi)\Gamma(-{\bar\nu})}\over{\pi}}\Big]r^{{1\over2}+{\bar\nu}}
-i{{\Gamma({\bar\nu})}
\over{\pi({\bar\omega}/2)^{{\bar\nu}}}}r^{{1\over2}-{\bar\nu}}\Big\}\
\end{eqnarray}
for the radial eigenfunctions which characterize the
Schr\"odinger differential equation (\ref{Eq9}) in the intermediate radial region
\begin{equation}\label{Eq26}
r_0\ll r \ll 1/\omega\  .
\end{equation}

\section{The discrete energy spectrum of the highly-excited bound-state resonances}

In the previous section we have explicitly shown that, for small resonant eigen-frequencies in the regime $\omega r_0\ll1$ \cite{Notewb}, there is an overlap radial region $r_0\ll r \ll 1/\omega$ [see Eqs. (\ref{Eq19}) and (\ref{Eq26})] in which the characteristic radial eigenfunctions $\psi(r)$ of the Schr\"odinger differential equation (\ref{Eq9}) are described by the two analytically derived mathematical expressions (\ref{Eq13}), (\ref{Eq16}), (\ref{Eq22}), and (\ref{Eq24}). In particular, inspecting the mathematical solutions (\ref{Eq18}) and (\ref{Eq25}) in the overlap region $r_0\ll r \ll 1/\omega$ (where {\it both} solutions are valid), one immediately realizes that these two expressions for the radial eigenfunctions $\psi(r)$ are characterized by the same functional (power-law) behavior. Matching the coefficients of the power-law $r^{{1\over2}+i\nu}$ terms in (\ref{Eq18}) and (\ref{Eq25}), one finds the dimensionless ratio \cite{Notegg}
\begin{equation}\label{Eq27}
{{B_1}\over{A_1}}={{{\bar\nu}\Gamma({\bar\nu})\Gamma\big({{2{\bar\nu}}\over{n-2}}\big)}
\over{i\pi[1+i\cot({\bar\nu}\pi)]\big({{i}\over{n-2}}\big)^{{2{\bar\nu}}\over{n-2}}({\bar\omega}
r_0/2)^{{\bar\nu}}}}\  .
\end{equation}
Likewise, matching the coefficients of the power-law $r^{{1\over2}-i\nu}$ terms in (\ref{Eq18}) and (\ref{Eq25}), one finds the dimensionless ratio \cite{Notegg}
\begin{equation}\label{Eq28}
{{B_1}\over{A_1}}={{i\pi\big[1+i\cot\big({{2{\bar\nu}}\over{n-2}}\pi\big)\big]\big({{i}
\over{n-2}}\big)^{{2{\bar\nu}}\over{n-2}}({\bar\omega}
r_0/2)^{{\bar\nu}}}\over{\big({{2{\bar\nu}}\over{n-2}}\big)\Gamma({\bar\nu})\Gamma\big({{2{\bar\nu}}\over{n-2}}\big)}}\
.
\end{equation}

Taking cognizance of Eqs. (\ref{Eq27}) and (\ref{Eq28}), one finds the characteristic resonance equation
\begin{equation}\label{Eq29}
({\bar\omega}
r_0/2)^{2{\bar\nu}}={{2}\over{(n-2)[1+i\cot({\bar\nu}\pi)][1+i\cot({{2{\bar\nu}}
\over{n-2}}\pi)]}}\Bigg[{{\nu\Gamma({\bar\nu})\Gamma\big({{2{\bar\nu}}\over{n-2}}\big)}\over{\pi\big({{i}
\over{n-2}}\big)^{{2{\bar\nu}}\over{n-2}}}}\Bigg]^2\  .
\end{equation}
It is important to emphasize again that the analytically derived resonance equation (\ref{Eq29}) for the characteristic binding energies (resonant frequencies) of the effective long-range Lennard-Jones potentials (\ref{Eq2}) \cite{NoteLR2} is valid in the small frequency regime [see Eqs. (\ref{Eq19}) and (\ref{Eq26})]\cite{Noterov}
\begin{equation}\label{Eq30}
\omega r_0\ll1\  .
\end{equation}

From Eq. (\ref{Eq29}) we finally obtain the discrete resonance spectrum \cite{Notect}
\begin{equation}\label{Eq31}
\omega r_0=2(n-2)^{{2}\over{n-2}}\Bigg\{{{2e^{{\pi\nu
n}\over{n-2}}}\over{(n-2)[1+\coth(\nu\pi)]
[1+\coth({{2\nu}\over{n-2}}\pi)]}}\Bigg[{{\nu\Gamma({\bar\nu})\Gamma\big({{2{\bar\nu}}\over{n-2}}\big)}
\over{\pi}}\Bigg]^2\Bigg\}^{{1}\over{2{\bar\nu}}} \times e^{-{{\pi
k}\over{\nu}}}\ \ \ ; \ \ \ k\in\mathbb{Z}\
\end{equation}
which characterizes the weakly bound-state (highly-excited) resonances of
the radial Schr\"odinger equation (\ref{Eq3}) with the effective long-range binding potentials (\ref{Eq2}). It is worth pointing out that the dimensionless expression on the r.h.s of Eq. (\ref{Eq31}) can be expressed in a more compact form. In particular, denoting
\begin{equation}\label{Eq32}
\phi_1\equiv\text{arg}[\Gamma({\bar\nu})]\ \ \ ; \ \ \
\phi_2\equiv\text{arg}\big[\Gamma\big({{2{\bar\nu}}\over{n-2}}\big)\big]\
,
\end{equation}
and using the identity (see Eq. 6.1.43 of \cite{Abram})
\begin{equation}\label{Eq33}
\Re[\Gamma(ix)]=\Big[{{\pi}\over{x\sinh(\pi x)}}\Big]^{1\over2}\  ,
\end{equation}
one can express the dimensionless resonance spectrum (\ref{Eq31}) in the remarkably compact form
\begin{equation}\label{Eq34}
\omega r_0=2(n-2)^{{2}\over{n-2}}e^{{\phi_1+\phi_2}\over{\nu}}\times e^{-{{\pi k}\over{\nu}}}\ \ \ ; \ \ \ k\in\mathbb{Z}\  .
\end{equation}

\section{Summary}

The discrete energy spectra of composite {\it long}-range
Lennard-Jones potentials of the form (\ref{Eq2}) were studied {\it
analytically}. Interestingly, using a functional matching procedure
for the characteristic eigenfunctions of the radial Schr\"odinger
equation (\ref{Eq9}), we have derived a remarkably compact
analytical formula for the dimensionless binding energies
\begin{equation}\label{Eq35}
{\cal E}(\nu,n;k)\equiv\Big({{2\mu\beta^{{{2}\over{n}}}}\over{\hbar^2}}\Big)^{{{n}\over{n-2}}}\cdot E(\nu,n;k)
\end{equation}
which characterize the highly-excited bound-state resonances of the long-range Lennard-Jones potentials (\ref{Eq2}) \cite{NoteLR2} [see Eqs. (\ref{Eq5}), (\ref{Eq7}), (\ref{Eq32}), and (\ref{Eq34})]:
\begin{equation}\label{Eq36}
{\cal E}(\nu,n;k)=-\big[2(n-2)^{{2}\over{n-2}}e^{{\phi}\over{\nu}}\big]^2\times e^{-{{2\pi k}\over{\nu}}}\ \ \ ; \ \ \ k\in\mathbb{Z}\  ,
\end{equation}
where
\begin{equation}\label{Eq37}
\phi\equiv\text{arg}[\Gamma({\bar\nu})]+\text{arg}\big[\Gamma\big({{2{\bar\nu}}\over{n-2}}\big)\big]\
.
\end{equation}

Finally, it is worth noting that the long-range binding potentials
(\ref{Eq2}) studied in the present paper are characterized by an
attractive inverse square law interaction term which naturally
appears in diverse quantum systems (see
\cite{Sq1,Sq2,Sq3,Sq4,Sq5,Sq6,Sq7,Sq8,Sq9,Sq10} and references
therein). Thus, our results may be of practical importance in the
physics of polarized molecules \cite{Sq3,Sq4,Sq5,Sq6} and in the
physics of composite polymers \cite{Sq9,Sq10} where this long-range
interaction term naturally appears. In addition, the famous Efimov
effect \cite{Sq7,Sq8} is characterized by the inverse square law
potential term, and thus our results may also be of practical
importance for physical models describing the quantum interactions
of bosonic particles \cite{Nis}.

\bigskip
\noindent
{\bf ACKNOWLEDGMENTS}
\bigskip

This research is supported by the Carmel Science Foundation. I thank
Yael Oren, Arbel M. Ongo, Ayelet B. Lata, and Alona B. Tea for stimulating discussions.



\begin{thebibliography}{99}
\bibitem{NotePa} The repulsive interaction term in the Lennard-Jones potential (\ref{Eq1}) is usually attributed in the literature to the quantum effect of Pauli repulsion at short distances.
\bibitem{Notekr} Here $k$ is the resonance parameter of the bound-state resonance [see Eq. (\ref{Eq31}) below].
\bibitem{Notelll} It is important to emphasize that
the physical case studied in the present work refers to eigenstates
with square angular number $l$ less than a critical value,
determined by $\alpha$, above which the total potential, which
vanishes asymptotically as $1/r^2$, becomes repulsive due to the
repulsive centrifugal force.
\bibitem{Notespt} It is worth noting that one learns from the
effective binding potential (\ref{Eq2}) that the effective radius
${\bar r}$ of the radial eigen-functions can be estimated by the
functional expression ${\bar r}=n\beta/2\alpha$. It should be
emphasized that, for physical systems which are made of molecular
gases, our analytical treatment is valid provided the effective
radius is less than the average half distance between the molecules.
\bibitem{Notew0} We shall assume, without loss of generality, that $\omega>0$.
\bibitem{Notese} It is worth noting that the small frequency regime (\ref{Eq8}) that we shall explore in the present paper corresponds to small binding energies in the regime $(2\mu\beta^{2/n}/\hbar^2)^{n/(n-2)}E\ll1$ [see Eqs. (\ref{Eq5}) and (\ref{Eq7})].
\bibitem{Notenu0} We shall assume, without loss of generality, that $\nu>0$.
\bibitem{Noterc} We recall that $0<\omega\in \mathbb{R}$ [see Eq. (\ref{Eq5}) and \cite{Notew0}]. One therefore finds the physically acceptable asymptotic spatial behavior $\psi(r\to\infty)\sim e^{-\omega r}\to 0$ for $B_2=0$.
\bibitem{Notewb} Note that this frequency regime corresponds to weakly bound-state (highly-excited) resonances of the
radial Schr\"odinger equation (\ref{Eq9}).
\bibitem{Notegg} Here we have used the functional identity
$-x\Gamma(x)\Gamma(-x)=\pi/\sin(\pi x)$ (see Eq. 6.1.17 of \cite{Abram}) in Eqs. (\ref{Eq18}) and (\ref{Eq25}).
\bibitem{Noterov} Note, in particular, that the two strong inequalities $r_0\ll r \ll 1/\omega$ in Eqs. (\ref{Eq19}) and (\ref{Eq26})
roughly correspond to an order-of-magnitude
difference ($r_0/r\lesssim 10^{-1}$ and $\omega r\lesssim 10^{-1}$) between the relevant physical parameters. One therefore realizes that the analytically derived equation (\ref{Eq29}) for the characteristic resonant frequencies of the effective long-range Lennard-Jones potentials (\ref{Eq2}) \cite{NoteLR2} is expected to be valid in the regime $\omega r_0\lesssim 10^{-2}$ of small binding energies.
\bibitem{Notect} Here we have used the relation $1=e^{i2\pi k}$, where the resonance parameter $k$ is an integer. In addition, we have used here the functional identity $i\cot(ix)=\coth(x)$.
\end{thebibliography}
\end{document}